\documentclass[10pt,a4paper,twoside]{article}
\usepackage{epsfig}
\usepackage{baltlat6}
\usepackage{array}
\usepackage{here}
\usepackage{graphics}
\usepackage{wrapfig}
\usepackage{floatfig}
\usepackage{graphicx}
\begin{document}
\ \
\vspace{0.5mm}
\setcounter{page}{1}
\vspace{8mm}

\titlehead{Baltic Astronomy, vol.\,21, 1--7, 2012}

\titleb{MODELING THE KINEMATICS OF DISTANT GALAXIES}

\begin{authorl}
\authorb{Rain Kipper}{1,2},
\authorb{Elmo Tempel}{2,3} and 
\authorb{Antti Tamm}{2}
\end{authorl}

\begin{addressl}
\addressb{1}{Institute of Physics, University of Tartu, 51010 Tartu, Estonia; rain@aai.ee}
\addressb{2}{Tartu Observatory, 61602 T\~oravere, Estonia}
\addressb{3}{National Institute of Chemical Physics and Biophysics, Tallinn, Estonia}
\end{addressl}

\submitb{Received: 2012 August 9; accepted: 2012 September 14}
\begin{summary} 
Evolution of galaxies is one of the most actual topics in astro-
physics. Among the most important factors determining the evolution are two
galactic components which are difficult or even impossible to detect optically:
the gaseous disks and the dark matter halo. We use deep Hubble Space Telescope images to construct a two-component (bulge + disk) model for stellar
matter distribution of galaxies. Properties of the galactic components are derived using a three-dimensional galaxy modeling software, which also estimates
disk thickness and inclination angle. We add a gas disk and a dark matter halo
and use hydrodynamical equations to calculate gas rotation and dispersion profiles in the resultant gravitational potential. We compare the kinematic profiles
with the Team Keck Redshift Survey observations. In this pilot study, two
galaxies are analyzed deriving parameters for their stellar components; both
galaxies are found to be disk-dominated. Using the kinematical model, the gas
mass and stellar mass ratio in the disk are estimated.

\end{summary}

\resthead{Modeling the kinematics of distant galaxies}{R. Kipper, E. Tempel, A. Tamm}

\sectionb{1}{INTRODUCTION}

Understanding the evolution of galactic structures is of uttermost importance
in astrophysics. The knowledge of the processes determining the morphology, star
formation history and intrinsic kinematics of galaxies give us the clues about the
general cosmological framework: fundamental cosmological parameters, properties
of dark matter, cosmic recycling and enrichment of baryonic matter, etc. However,
several key aspects of galaxy evolution are poorly understood so far.

%
Owing to extensive surveys, like the Sloan Digital Sky Survey (York et al.
2000), the general structure of galaxies is quite well established (e.g., Simard et
al. 2011; Lackner \& Gunn 2012). For a direct tracing of evolutionary effects, the
local galaxy sample has to be compared to galaxies at cosmologically significant
distances. Unfortunately, only small samples of distant galaxies have been studied
so far. From photometric decomposition of galaxies out to redshift $z\simeq6$ (e.g.,
Tamm \& Tenjes 2006; Fathi et al. 2012), it has been shown that sizes of galaxy
disks decrease with the redshift, in general accordance with simulations.

On the other hand, studies of the evolution of the intrinsic kinematics (and
thus the gravitation potential and dynamical formation history) of galaxies do not
extend much beyond $z\simeq1$. It has been shown that the evolution of the Tully-
Fisher relation is mild or missing over this period (Fern\'andez Lorenzo et al. 2010;
Miller et al. 2011).

In more detailed studies the photometric and kinematic data are used self-consistently to split the structure of galaxies into contributions by individual stellar, gaseous and dark matter components (Tenjes et al. 1994, 1998; Tempel \&
Tenjes 2006; Chemin et al . 2011; Jardel et al. 2011). However, for higher redshift
objects the observational data rarely provide possibilities for such analyses (Tamm
\& Tenjes 2003, 2005).

In the present study we have used both photometric and kinematic observations
to study the structure of distant disk-dominated galaxies. We have applied a simple
bulge\, +\, disk model on deep Hubble Space Telescope observations. Using the data
for gas kinematics, a gas disk component was added to the model. Masses of the
components were derived by solving the isotropic Jeans equations.

Throughout this paper we assume a Friedmann-Robertson-Walker cosmological model with the total matter density $\Omega_\mathrm{m}=0.27$, dark energy density $\Omega_\Lambda = 0.73$, and the Hubble constant $H_0 = 71\, \mathrm{km}\,\mathrm{s}^{-1}\mathrm{Mpc}^{-1}$.

\sectionb{2}{DATA}

\begin{wrapfigure}[13]{i}[0pt]{70mm}
\begin{tabular}{cc}
\psfig{figure=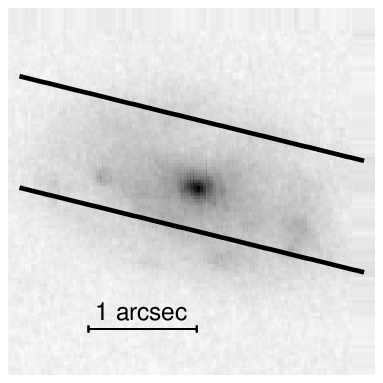,width=35mm,clip= 0 0 15}
\psfig{figure=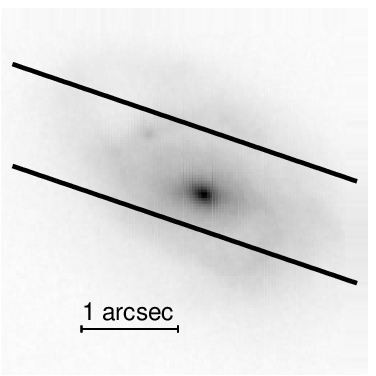,width=35mm,clip= 0 0 15}
\end{tabular}
\vspace{2mm}
\captionb{1}{Photometric images of two galaxies  (left TKRS 2568, right TKRS 9059)  in $z$ filter, with the position of the spectrograph slit overplotted.}
\end{wrapfigure} 

Photometry of galaxies is based on the deep Hubble Space Telescope ACS camera observations in the GOODS project (Giavalisco et al. 2004). Modelling is based on $z$ filter observations, in which young stars and dust have the smallest effect. Initial data reduction, including dithering (Mutchler \& Cox 2001) to 0.03\,arcsec/pixel had been done by the GOODS team. Intensity counts were transformed into magnitudes using the formula
\begin{equation}
m=-2.5\log_{10}(N)+ZP,
\label{eq_N_to_m}
\end{equation}
where $N$ is electron counts per second in CCD pixel and $ZP$ is zero point of the used filter. According to Sirianni et al. (2005), the zero point for the $z$ filter is 24.86 mag. We took 4.51 as the value of solar luminosity.

Figure~1 shows the $z$ filter images of the studied galaxies, with the spectrograph
slit positions and sizes indicated. Both galaxies have been selected to be disk-
dominated objects with relatively good spectroscopic observations.

We used spectroscopic observations from the Team Keck Redshift Survey (TKRS;
Wirth et al. 2004) of galaxies in the GOODS field using the DEIMOS spectro-
graph of the Keck II telescope. They have measured the [O II] 3727\,\AA\/ emission line
along the slit roughly parallel to the major axes of galaxies. The slit width was
$1\arcsec$, which is comparable with the galaxy size. During spectroscopic observations the seeing was $0.6\arcsec$ to $1\arcsec$.

The original sample contains the rotation curves and velocity dispersion profiles
of 380 galaxies extracted by Weiner et al. (2006) from the TKRS survey data.
Since edge-on galaxies are difficult to use for photometric analysis and the face-on
ones for kinematic analysis, we chose galaxies with intermediate inclination angles.
Additional selection was based on the symmetry, spatial and velocity extent of
the observed kinematics, required for reliable modeling. Approximately 10\% of
the galaxies matched these criteria. In this study we present the analysis of two
galaxies (TKRS 2568 and TKRS 9059) which have good rotation and dispersion
profiles and have also regular and disk-dominated morphology. The nomenclature
is the same as used by Weiner et al. (2006).

\sectionb{3}{MODEL}
We used a three-dimensional model to describe the structure of each galaxy, briefly reviewed below. Details of the general model are given in Tempel et al. (2010, 2011).

\subsectionb{3.1}{PHOTOMETRIC MODEL}
The model galaxy is given as a superposition of its individual stellar components. In the present study we used two component model: bulge\,+\,disc. The light distribution of both components are approximated by the Einasto law,
\begin{equation}
l(a)=l(0)\exp\left[-\left({a}/{ka_0}\right)^{1/N}\right],
\label{eq_Einasto}
\end{equation} 
where $l(0)=hL/(4\pi qa_0^3)$ is the central density and $L$ is the component luminosity; $a_0$ is the mean harmonic radius and $a=\sqrt{R^2+z^2/q}$, where $R$ and $z$ are cylindrical coordinates and $q$ is axial ratio of a component. The coefficients $h$ and $k$ are normalising parameters, dependent on $N$.

These components are projected onto the plane of the sky using the following formula: 
\begin{equation}
L(X,Y) = 2\sum_{j} \frac{q_j}{Q_j}\int_A^\infty \frac{l_j(a)a\textrm{d}a}{\sqrt{a^2-A^2}},
\label{eq_vaatejoon}
\end{equation}
where $Q^2=\cos^2i+q^2\sin^2i$ is the apparent axial ratio, $i$ is the inclination angle of a galaxy, 
$A=\sqrt{X^2+Y^2/Q^2}$ is the major semi-axis of the equidensity ellipse of the projected light distribution. The summation is made over galaxy components.

We estimated the correctness of the model fit, using the $\chi^2$ value, defined by the sum of the squared differences of the model and observational pictures. 

For minimizing $\chi^2$ we gave a physically justified initial guess for the parameters
ourselves. In the next step, the parameter values were specified using the downhill
simplex method of Nelder and Mead from the Numerical Recipe library.

The photometric model gave us the stellar component (bulge and disc) parameters ($a_0$, $q$, $N$, $L$) and the inclination angle of a galaxy. These parameters were kept fixed during the following dynamical modelling.

\subsectionb{3.2}{DYNAMICAL MODEL}
The dynamical model is based on the density distributions of the stellar components (bulge and disk), the gas disk and the dark matter halo. Density profiles of the stellar components have the same distribution as the luminous components;
a constant mass-to-light ratio was assumed, different for bulge and disk. The gas
disk and dark matter parameters are derived during dynamical modeling. For
describing the gas disk density distribution, we used Eq.~(\ref{eq_Einasto}). For the dark matter distribution, we used the Einasto profile in another form, which is becoming
increasingly popular for this purpose:
\begin{equation}
    \rho_\mathrm{Einasto}=\rho_\mathrm{c}\exp\left\{-d_n\left[\left( {r}/{r_\mathrm{c}}\right)^{1/n} -1\right]\right\},
\end{equation}
where $n$ is in principle a free parameter. According to $N$-body simulations, we take $n=6.0$ (Merritt et al. 2006; Navarro et al. 2010). The term $d_n$ is a function of $n$ in a way that $\rho_\mathrm{c}$ is the density at $r_\mathrm{c}$ defining a half-mass radius. The value of $d_n$ is 17.67 for $n=6$ (Merritt et al. 2006).

Dynamics of a galaxy was calculated from the potential and density distributions using the Jeans equations. For simplicity, axisymmetry and stationarity of
the galaxy were assumed for solving the Jeans equations. Since the observed dynamics is based on gas motions, we could use the isotropic Jeans equations. The
Jeans equations were thus used in the form
\begin{equation}
\frac{\partial(\rho\sigma^2)}{\partial R} + \frac{\rho V^2_\varphi}{R} + \rho\frac{\partial\Phi}{\partial R} = 0, \qquad
\frac{\partial(\rho\sigma^2)}{\partial z} + \rho\frac{\partial\Phi}{\partial z} = 0,
\end{equation}
where $\Phi$ is the sum of potentials of galaxy components, $V_\varphi$ is rotational velocity, $\sigma$ is dispersion of velocities and $\rho$ is gas disc density distribution; $R$ and $z$ are cylindrical coordinates. 

From these equations rotational velocity and velocity dispersions were derived
\begin{eqnarray}
V_\varphi^2(R,z)&=&\frac{-R}{\rho}\left[\frac{\partial(\int_z^\infty\rho\frac{\partial\Phi}{\partial z}\textrm{d}z')}{\partial R}+\rho\frac{\partial\Phi}{\partial R}\right],
\label{eq_vrot}
\\
\sigma^2(R,z)&=&\rho^{-1}\int_z^\infty\rho(R,z')\frac{\partial\Phi(R,z')}{\partial z}\textrm{d}z'.
\label{eq_disp}
\end{eqnarray}

To make the modeled data comparable with observations, we needed to find
the line-of-sight projected velocity distributions. For constructing the line-of-sight
distribution we applied a Gaussian velocity profile (based on $V_\varphi$ and $\sigma$) in every
point along the line of sight, and integrated them using the square of gas density
as a weight. The projected velocity profile $F_{(X,Y)}(V)$ corresponding to each line
of sight ($X,Y$) was found using the equation

\begin{equation} F_{(X,Y)}(V)=\int\limits^{\infty}_{X}
  \frac{\sum\limits_{j=1}^2\left[l^2_\mathrm{gas}(R,z_j)\,f_{(R,z)}(V)\right]}{\sin{i}\,\sqrt{R^2-X^2}}R\,\mathrm{d}R,\label{eq:los_profile}
\end{equation}
\begin{equation} z_{1,2}= \frac{Y}{\sin{i}} \pm \frac{\sqrt{R^2-X^2}}{\tan{i}},
\end{equation}
where $l_\mathrm{gas}(R,z)$ denotes the spatial gas density (given by Eq.~(\ref{eq_Einasto})) and $f_{(R,z)}(V)$ is the corresponding line-of-sight Gaussian velocity profile at a given point ($R,z$) in the galaxy, calculated on the basis of Eqs.~(\ref{eq_vrot}) and (\ref{eq_disp}).

\subsectionb{3.3}{COMPARING THE VELOCITY PROFILE WITH OBSERVATIONS}

The observational data points of velocity dispersion and rotation are given
along the spectroscopic slit. To compare these data with our model, we had to
take into account the width of the slit and seeing. To achieve this we firstly
calculated the model output as a pixel map, where each pixel represents a line-of-
sight integrated velocity distribution profile.

The atmospheric effects were taken into account by convolution of the velocity
profile map with a Gaussian seeing function. During the observations the seeing
changed from 0.6\arcsec\/ to 1\arcsec\/ , therefore we used a Gaussian seeing kernel with a width
of 0.8\arcsec . Since the spectroscopic slit was of similar width (1\arcsec ), the accuracy of the
seeing kernel was not very important.

%
Trying to take into account the slit width, we summed the velocity profiles
along the slit in bins, centered to the given observational points. We used the
square of spatial density of gas as a weight for the summation. Thus the final
model velocity profile allows a direct comparison to the observational points.

For fitting the model galaxy to the observed kinematics, we used the structure
of the bulge and disk as derived during the photometric analysis. We used the
maximal disk method, ascribing as much mass to the disk as allowed by the observed rotation curve. Subsequently, the masses of the bulge, gas disk and dark
matter halo were derived to achieve the best fit to the observed kinematics.

\begin{table}[!t]
\begin{center}
\vbox{\small\tabcolsep=6pt
\parbox[c]{124mm}{\baselineskip=10pt
{\normbf\ \ Table 1. }{Parameters of galaxy models.}
\label{tab_phot}}
\begin{tabular}{lcc|lccccc}
\hline
\hline
ID & $z$  & $i$ & Comp & $a_0$ & $q$ & $N$ & $L_\mathrm{total}$ & $M$ \hstrut\\ 
  & & [deg] & & [kpc] & & & $[10^{10}L_\odot]$ & $[10^{10}M_\odot]$\\
\hline
& & & Bulge & 0.9 & 1.0 & 5.1 & 2.4 & 7.0\\
2568 & 0.488 & 49 & Stellar disc & 6.68 & 0.2 & 0.5 & 22.3 & 7.0 \\
 & & & Gas disc & 8.0 & 0.1 & 0.1 &  & 1.0\\
 \hline
 & & & Bulge & 1.2 & 1.0 & 6.0 & 3.7 & 4.8\\
 9059 & 0.253 & 65 & Stellar disc & 5.95 & 0.2 & 0.7 & 19.1 & 7.0 \\
 & & & Gas disc & 10.0 & 0.1 & 0.2 &  & 3.0\\
\hline
\end{tabular}
}
\end{center}
\vskip-4mm
\end{table}

\sectionb{4}{RESULTS} 

Using the IRAF task \emph{ellipse}, we found elliptically averaged one-dimensional
profiles of the model and the observed images using the same ellipse parameters
for both cases. Figure~2 shows the model and the observed profiles in the $z$ filter,
together with the contributions by the bulge and disk components. The modeled
galaxy parameters are given in Table~1. The results of the dynamical modeling
are presented in Figure~3, showing the observed kinematic profiles together with
the modeled ones.

\begin{figure}[!tH]
\vbox{
\centerline{
\psfig{figure=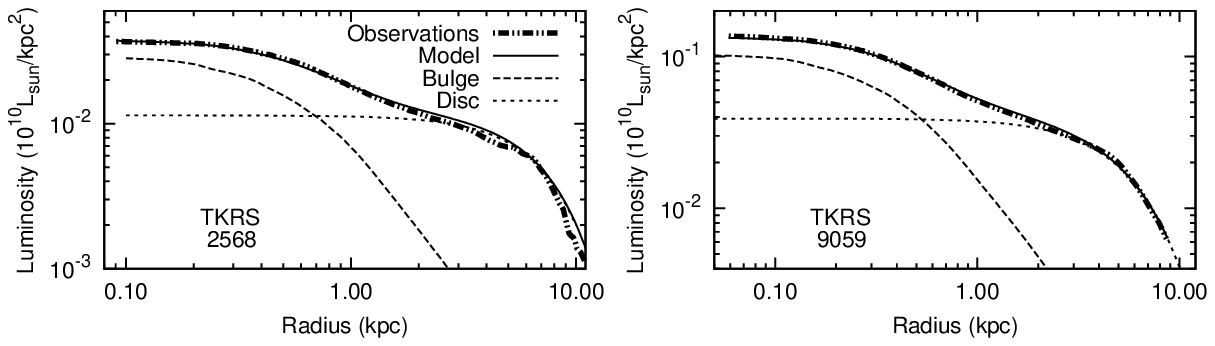,width=124mm,angle=0,clip=}
}
\captionb{2}
{Results of photometric modelling for the $z$ filter. }
\label{fit_phot_profiles}
}
\end{figure}
\begin{figure}[!tH]
\vbox{
\vspace{-5mm}
\centerline{
\psfig{figure=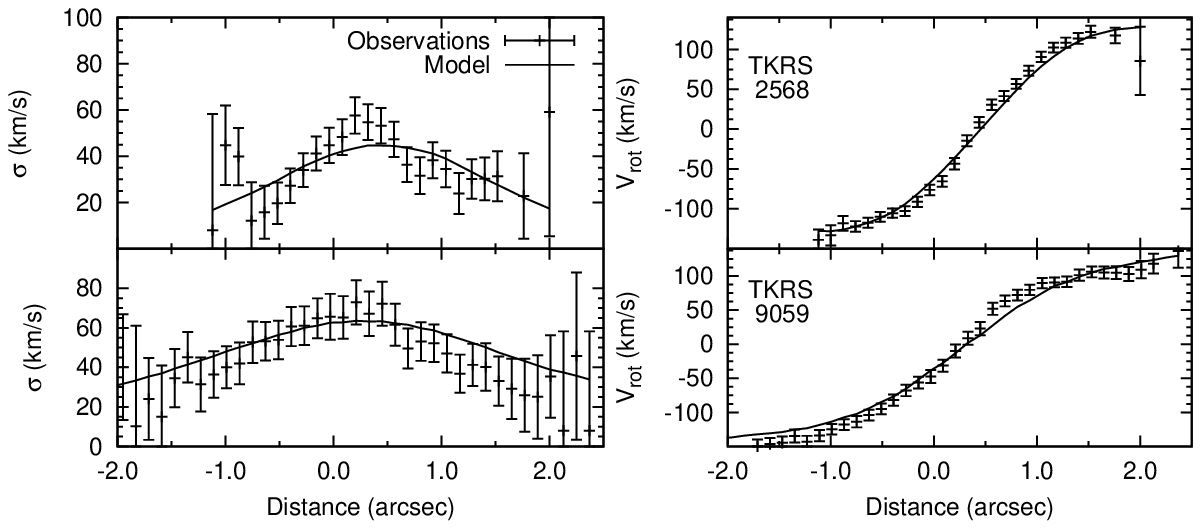,width=124mm,angle=0,clip=}
}
\vspace{1mm}
\captionb{3}
{Results of dynamical modelling for: TKRS 2568 and TKRS 9059. The x axis is from the slit center.}
\label{fig_dyn_profiles}
}
\end{figure}

Due to degeneracy, the found parameters are one of the possible parameter sets
that fit the galaxy. For this the parameters of the dark matter halo are responsible,
because they originate from the inner parts of the galaxy, where the potential is
dominated by the luminous components.

TKRS\,2568 is a disk-dominated (bulge to total luminosity ratio B/T = 0.1)
galaxy at redshift $z$ = 0.488. The inclination angle is $49^\circ$ , which makes the galaxy
suitable for kinematic and photometric analysis. It has clumpy disk structure,
which could refer to a forming bulge (Conselice 2003). As can be seen from Figures~2 and~3, the four-component model (bulge + disk + gas + dark matter halo) gives
a good fit to the photometric and kinematic observations. Since the observed
rotation curve does not reach the expected plateau supported by the gravitational
potential of dark matter halo, the parameters for the dark matter component are
highly uncertain.

TKRS 9059 is also disk-dominated galaxy (B/T = 0.14) at redshift 0.253. In
the outer regions of the galaxy stellar streams or tilted spiral arms, maybe be re-
lated to a recent merger or accretion event, are seen. They make the interpretation
of the kinematic data less reliable: stationarity might not be fully achieved, which
is also seen from higher dispersion errors. The inclination angle ($67^\circ$) is determined from the inner, expected disk-dominated region. A good fit to the observed
kinematics does not require the use of a dark matter halo component. Although
not justified by the current cosmological paradigm, the data do not allow us to
give any constraints to the possible dark matter halo.

In general, resulting from the model constraints and the limited depth of observations, the derived galactic components are somewhat degenerate and the presence of the dark matter halos remain not confirmed. However, the modeling software enables to brake the degeneracy between the thickness and the inclination
angle of stellar disk, and a corresponding study of larger samples would be important for detection of evolutionary effects in disk properties.

%

\thanks{
This study is based on observations made with the
NASA/ESA Hubble Space Telescope, obtained from the data archive at the Space
Telescope Science Institute. Our work was supported by the Estonian Science
Foundation grants 7765, 8005, 9428, MJD272 and the projects SF0060067s08 and
TK120 in (Astro)particle Physics and Cosmology (TK120). All the figures have
been made using the Gnuplot plotting utility. We thank the TKRS group and
B. Weiner for making their data publicly available and the anonymous referee for
useful comments and suggestions.
}

\References
\refb Chemin L., de Blok W. J. G., Mamon G. A. 2011, AJ, 142, 109
\refb Conselice C. J. 2003, ApJS, 147,1
\refb Fathi K., Gatchell M., Hatziminaoglou E., Benoit E. 2012, MNRAS, 423, L112
\refb Fern\'andez Lorenzo M., Cepa J., Bongiovanni A. et al. 2010, A\&A, 521, A27
\refb Giavalisco M., Ferguson H.C., Koekemoer A.M., et al. 2004, AJL, 600, L93
\refb Jardel J.R., Gebhardt K., Shen J. et al. 2011, ApJ, 739, 21
\refb Merritt D., Graham A. W., Moore B. et al. 2006, AJ, 132, 2685
\refb Miller S.H., Bundy K., Sullivan M., et al. 2011, ApJ, 741, 115
\refb Lackner C. N., Gunn J. E. 2012, MNRAS, 421, 2277
\refb Mutchler M., Cox C. 2001, Instrument Science Report ACS
\refb Navarro J. F., Ludlow A., Springel V. et al. 2010, MNRAS, 402, 21
\refb Simard L., Mendel J.T., Patton D. R. et al. 2011, APJS, 196, 11
\refb Sirianni M., Jee M. J., Ben{\'{\i}}tez N. et al. 2005, ASP, 117, 1049
\refb Tamm A., Tenjes P., 2003, A\&A, 403, 529
\refb Tamm A., Tenjes P. 2005 A\&A 433, 31
\refb Tamm A., Tenjes P. 2006, A\&A, 449, 67
\refb Tempel E., Tenjes P., 2006, MNRAS, 371, 1269
\refb Tempel E., Tamm A., Tenjes P. 2010, A\&A, 509, A91
\refb Tempel E., Tuvikene T., Tamm A. et al. 2011, A\&A, 526, A155
\refb Tenjes P., Haud U., Einasto J., 1994, A\&A, 286, 753
\refb Tenjes P., Haud U., Einasto J., 1998, A\&A, 335, 449
\refb Weiner B.J. Willmer C.N.A. Faber S.M. et al. 2006, ApJ, 653, 1027
\refb Wirth G.D., Willmer C.N.A., Amico P. et al. 2004, AJ, 127, 3121
\refb York D.G., Adelman J., Anderson Jr. J.E., et al. 2000, AJ, 120, 1579
\end{document}